\documentclass[aps,prb,amsmath,twocolumn,amssymb,floatfix,nosuperscriptaddress,nofootinbib]{revtex4-2}

\usepackage{MnSymbol}
\usepackage{braket}
\usepackage[dvipsnames]{xcolor}
\usepackage{float}
\usepackage{subfigure}
\usepackage{tikz}
\usepackage[colorlinks=true,linktoc=page,linkcolor=red,citecolor=blue,urlcolor=purple]{hyperref}
\usepackage{mathtools}

\usepackage{amsfonts}

\usepackage{pifont}

\usepackage{enumerate}

\mathchardef\mhyphen="2D 

\newcommand\bea{\begin{eqnarray}}
\newcommand\eea{\end{eqnarray}}
\newcommand\beq{\begin{equation}}  
\newcommand\eeq{\end{equation}}

\newcommand{\vb}{\mathbf}
\usepackage[normalem]{ulem}
\definecolor{lime}{HTML}{A6CE39}
\usepackage{sidecap,tikz}
\DeclareRobustCommand{\orcidicon}{\hspace{-1.0mm}
	\begin{tikzpicture}
		\draw[lime, fill=lime] (0.0,0.0) 
		circle [radius=0.15] 
		node[white] {{\fontfamily{qag}\selectfont \tiny \,ID}};
		\draw[white, fill=white] (-0.0525,0.095) 
		circle [radius=0.007];
	\end{tikzpicture}
	\hspace{-3.0mm}
}
\foreach \x in {A, ..., Z}{\expandafter\xdef\csname orcid\x\endcsname{\noexpand\href{https://orcid.org/\csname orcidauthor\x\endcsname}{\noexpand\orcidicon}}
}

\usepackage{filecontents}

\begin{document}


\title{Interplay between altermagnetism and topological superconductivity\\  on an unconventional superconducting platform}  

\author{Pritam Chatterjee\orcidA{}}
\email{pritam@nt.phys.s.u-tokyo.ac.jp}
\affiliation{Department of Physics, The University of Tokyo, 7-3-1 Hongo, Bunkyo-ku, Tokyo 113-0033, Japan}
\affiliation{Institute of Physics, Sachivalaya Marg, Bhubaneswar-751005, India}
\affiliation{Homi Bhabha National Institute, Training School Complex, Anushakti Nagar, Mumbai 400094, India}

\author{Vladimir Juri\v{c}i\'c\orcidB{}}
\email{vladimir.juricic@usm.cl}
\affiliation{Departamento de F\'isica, Universidad T\'ecnica Federico Santa Mar\'ia, Casilla 110, Valpara\'iso, Chile}

\begin{abstract}
 We propose a theoretical model to investigate the interplay between altermagnetism and $p$-wave superconductivity, with a particular focus on topological phase transitions in a two-dimensional (2D) $p$-wave superconductor, considering both chiral and helical phases. Our study reveals that the emergence of helical and chiral Majorana states can be tuned by the amplitude of a $d-$wave altermagnetic order parameter, with the outcome depending  on the nature of the superconducting state.  In the helical superconductor, such an altermagnet can  induce a topological phase transition into a gapless topological superconductor hosting Majorana flat edge modes. On the other hand, in the chiral superconductor, the topological transition takes place  between a topologically nontrivial gapped phase and a  gapless nodal-line superconductor, where the Bogoliubov quasiparticle bands intersect at an isolated line in momentum space. Remarkably,  we show that when such an  altermagnet is coupled to  a mixed-pairing superconductor, with both chiral and helical components, a hybrid topological phase emerges, featuring dispersive Majorana edge  modes that coexist with nearly flat Majorana edge states. Our findings therefore establish a novel platform for controlling and manipulating Majorana modes  in unconventional superconductors with vanishing total magnetization.           
\end{abstract}

\maketitle

\section{Introduction}

Altermagnetism, a recently established form of magnetic order, bridges the gap between ferromagnetism and antiferromagnetism~\cite{Sinova2022b,sinova2022,Turek2022}.  While exhibiting zero net magnetization, a characteristic often associated with antiferromagnets, it simultaneously induces spin-split electronic band structures—a hallmark of ferromagnetism. This unique feature  has sparked significant interest within both theoretical and experimental condensed matter communities, with works focusing on purely  altermagnetic features~\cite{Sinova2022b,sinova2022,Turek2022,McClarty2024,Sun2023b,Hodt2024,Das2023}, various superconductor responses to an altermagnetic state \cite{Mazin2022,Beenakker2023,ghorashi2023,Papaj2023,Sun2023,Li2023,Li2024,Zhu2023,Ouassou2023,Zhu2024,Zhang2024,Wei2024,Chakraborty2024,mondal2024,maeda2025-arXiv,Hong-PRB2025}, and  electronic topology in the presence of  altermagnetism~\cite{Fernandes2024,dasroy2024}. The intriguing nature of altermagnetism extends further since  its textures manifest as even-momentum higher-harmonic orders (e.g., $d-$, $g-$wave), thus providing the previously lacking magnetic analogs to unconventional superconducting (SC) orders, such as  $d-$, $g-$wave, etc.  This opens exciting avenues for exploring novel material properties and functionalities, which requires further understanding of the underlying mechanisms governing altermagnetism and its potential applications.




\begin{table*}[t!]
\centering
\begin{tabular}{|l|c|c|c|c|c|}
\hline
\text{System} & \text{PHS} & \text{TRS} & \text{Chiral Symmetry} & \text{Symmetry Class} & \text{Topological Invariant} \\
\hline
\hline
Helical SC & \ding{51} & \ding{51} & \ding{51} & DIII & $\mathbb{Z}_2$ \\
\hline
Chiral SC & \ding{51} & \ding{55} & \ding{55} & D & $\mathbb{Z}$ \\
\hline
Helical SC + Altermagnet & \ding{51} & \ding{55} & \ding{55} & D & $\mathbb{Z}$ \\
\hline
Chiral SC + Altermagnet & \ding{51} & \ding{55} & \ding{55} & D & $\mathbb{Z}$ \\
\hline
\end{tabular}
\caption{Topological classification of the Bogoliubov--de Gennes (BdG) Hamiltonian in Eq.~\eqref{Eq1} for different superconducting (SC) pairings, with and without coupling to the $d$-wave altermagnetic order parameter. Anti-unitary particle-hole symmetry (PHS) and time-reversal symmetry (TRS) operators square to $+1$ and $-1$, respectively, with the corresponding topological invariants determined according to the tenfold classification of topological insulators and superconductors~\cite{ryu2010topological}. The symbols \ding{51} and \ding{55} denote the presence and absence of the symmetry, respectively. See App.~\ref{app:symmetries} for details of the symmetry transformations.}
\label{tab:symmetries}
\end{table*}

The realization of one- and two-dimensional (2D) topological superconductors has been a key impetus  in the study of topological phases of matter, driven by the goal of creating localized Majorana zero modes, which are considered as fundamental building blocks for topological quantum computation~\cite{Chiu2016,Sato2017,Nayak2008,Lutchyn2010,Nadj-Perge2013,Qi2010,Saha2017,Chatterjee2023}. Furthermore, there has recently been growing interest in gapless topological superconductors from both theoretical and experimental perspectives~\cite{Wiesendanger2024,Chatterjee2024,Chatterjee2024b,Subhadarshini2024,Nakosai2013,Sedlmayr2015,Wang2017,Zhang2019,Chen2015}, which may host zero-energy Majorana flat edge modes 
(MFEMs) instead of isolated Majorana zero modes, with a very recent experiment reporting the signature of MFEMs in a magnetic-superconductor hybrid system composed of Fe/Ta(110)~\cite{Wiesendanger2024}. We here present an alternative  approach to introduce gapless topological superconductors  hosting MFEMs, along with other new topological SC phases, by invoking  an unconventional SC platform involving  an altermagnet, therefore carrying a net zero magnetization, that, to the best of our knowledge, has not been explored until now.

To this end, we here  consider the simplest $d$-wave altermagnetic order parameter (OP) in the presence of both chiral and helical $p$-wave SC states, as well as of the hybrid chiral-helical SC phase. We demonstrate that coupling such an altermagnetic  OP featuring  the spin oriented in the $x-$direction with this unconventional superconductor can give rise to several gapped and gapless topological SC phases. For a helical $p$-wave superconductor, altermagnetism drives the system from a helical gapped topological superconductor to a gapless topological superconductor that hosts MFEMs  (Figs.~\ref{Fig1} and~\ref{Fig3}). In contrast, for a chiral $p$-wave superconductor, a topological phase transition occurs  from a chiral gapped $p$-wave topological superconductor to a gapless nodal-line superconductor in the presence of altermagnetism (Figs.~\ref{Fig2} and~\ref{Fig4}). Notably, for a mixed $p$-wave superconductor, combining helical and chiral SC OPs, altermagnetism induces an unusual hybrid topological phase where both dispersive and flat Majorana edge modes coexist, as displayed in Fig.~\ref{Fig5}. In all the discussed cases, the topological phases are separated by the bulk band closing, as explicitly shown in Fig.~\ref{Fig6}. Finally, this work reveals a tunable altermagnet-unconventional-superconductor platform that can support both gapless topological superconductivity, and a hybrid topological SC phase that features  coexisting dispersive and flat Majorana edge modes. 

The rest of this paper is organized as follows. In Sec.~\ref{Sec:II}, we introduce the model for the 2D altermagnet with $d$-wave magnetic ordering coupled with  unconventional helical and chiral $p$-wave superconductors. Next, we present numerical results on the edge states using the nanoribbon termination in the square lattice geometry in Sec.~\ref{Sec:III}, with the results for the edge state spectrum and the local density of states (LDOS) shown in Sec.~\ref{subSec:I} and Sec.~\ref{subSec:II}, respectively, for the chiral and helical phases, while in Sec.~\ref{subSec:III} we discuss both the edge states and LDOS in the case of mixed pairing phase. In Sec.~\ref{Sec:IV}, we compute the bulk gap profile as a signature of a topological phase transition by analyzing  the corresponding lattice model. Finally, we summarize and discuss our results  in Sec.~\ref{Sec:V}. Analysis of symmetries of the SC Hamiltonian is presented in App.~\ref{app:symmetries}, while the details of the edge theory in purely helical SC phase are discussed in App.~\ref{sec:edge_helical}.

\section{Model and Method}\label{Sec:II}

In this work, we analyze a minimal model that captures the nontrivial interplay between altermagnetism and 2D unconventional superconductivity. Specifically, we consider a Bogoliubov–de Gennes (BdG) Hamiltonian that incorporates both helical and chiral $p$-wave SC orders in the presence of an altermagnetic background. The full BdG Hamiltonian describing this hybrid system reads
\begin{equation}
\label{Eq1}
H = \frac{1}{2} \sum_{\bf k} \Phi_{\bf k}^{\dagger} h_{\bf k} \Phi_{\bf k},
\end{equation}
where $\Phi_{\bf k} = \left( c_{{\bf k}\uparrow}, c_{{\bf k}\downarrow}, c_{-{\bf k}\uparrow}^{\dagger}, c_{-{\bf k}\downarrow}^{\dagger} \right)^\top$ is the four-component Nambu spinor. Here, $c_{{\bf k}\uparrow}$ annihilates an electron quasiparticle with momentum ${\bf k}$ and spin $\uparrow$, while $c_{-{\bf k}\downarrow}^\dagger$ creates a quasiparticle with opposite momentum and spin. The BdG kernel $h_{\bf k}$ takes the form 
\begin{eqnarray}
h_{\bf k} &=& \xi_{\bf k} \tau_z s_0 + J({\bf k}) \tau_0 s_x 
+ \Delta_p^h \left( \tau_y s_0 \sin k_x - \tau_x s_z \sin k_y \right) \nonumber \\
&& + \Delta_p^c \left( \tau_x s_0 \sin k_x - \tau_y s_0 \sin k_y \right),
\label{Eq2}
\end{eqnarray}
where $\tau_i$ and $s_i$ ($i=0,x,y,z$) are Pauli matrices acting in particle-hole (Nambu) and spin space, respectively, and the free-quasiparticle dispersion is given by
\begin{equation}
\xi_{\bf k} = -2t (\cos k_x + \cos k_y) + 4t - \mu,
\end{equation}
where $t$ is the nearest-neighbor hopping amplitude and $\mu$ is the chemical potential.

The term $\sim J({\bf k})$ encodes the coupling of the altermagnetic order to the electron quasiparticles, which we assume to possess  $d$-wave symmetry,
\begin{equation}\label{eq:altermagnet}
J({\bf k}) = 2J_A(\cos k_x - \cos k_y),
\end{equation}
where $J_A$ is the amplitude of the altermagnetic exchange field~\cite{Li2023,ghorashi2023,sinova2022,Šmejkal2022}. The momentum-dependent structure of the altermagnetic OP leads to spin-split quasiparticle bands that {break fourfold rotational symmetry} (\( C_4 \)), \( J({\bf k}) \rightarrow -J({\bf k}) \) under $C_4$.  In fact, this antisymmetry under lattice rotations distinguishes altermagnetism from conventional ferromagnetic exchange, which is momentum-independent. Moreover, despite inducing spin splitting, the altermagnetic order preserves {zero net magnetization} over the Brillouin zone, as the contributions from different momenta cancel out. The altemagnetic term also breaks time-reversal symmetry (TRS) explicitly, since under the time-reversal operation \( \mathcal{T} \), one has \( s_x \rightarrow -s_x \) while \( J({\bf k}) \) remains invariant, such that \( \mathcal{T}[J({\bf k}) s_x]\mathcal{T}^{-1} =- J({\bf k}) s_x \). 
This results in a nontrivial coupling to the SC sector, modifying the quasiparticle spectrum in a momentum-anisotropic and symmetry-dependent fashion, which, in turn, also pertains to the topological features of the Bogoliubov band structures in the resulting SC  state, as we show in the following.

The two SC OPs considered in this model correspond to different pairing symmetries. The helical $p$-wave OP, $\Delta_p^h$, features a spin-triplet pairing with a spin-dependent orbital structure preserving TRS, while the chiral $p$-wave OP, $\Delta_p^c$, breaks time-reversal, see also Table~\ref{tab:symmetries}.  A more detailed discussion on the symmetries is deferred to App.~\ref{app:symmetries}.

We now analyze the quasiparticle dispersion and the band structure of the superconductor-altermagnet system for both pairing symmetries. In the helical case, the energy spectrum takes the form
\begin{equation}
\epsilon_{\bf k} = r \sqrt{ \xi_{\bf k}^2 + J({\bf k})^2 - \frac{1}{2} (\Delta_p^h)^2 (\cos 2k_x - \cos 2k_y) + s F({\bf k}) },
\label{Eq3}
\end{equation}
where $r,s = \pm$ label the band and particle-hole indices, respectively, and
\begin{equation}
F({\bf k}) = 2J({\bf k}) \sqrt{ \xi_{\bf k}^2 + (\Delta_p^h \sin k_x)^2 }.
\end{equation}
The term \( F({\bf k}) \) captures the momentum-resolved coupling  between the helical pairing and the altermagnetic field, leading to anisotropic features in the spectrum that can drive topological transitions.

\begin{figure}[t!]
\centering
\subfigure{\includegraphics[width=0.48\textwidth]{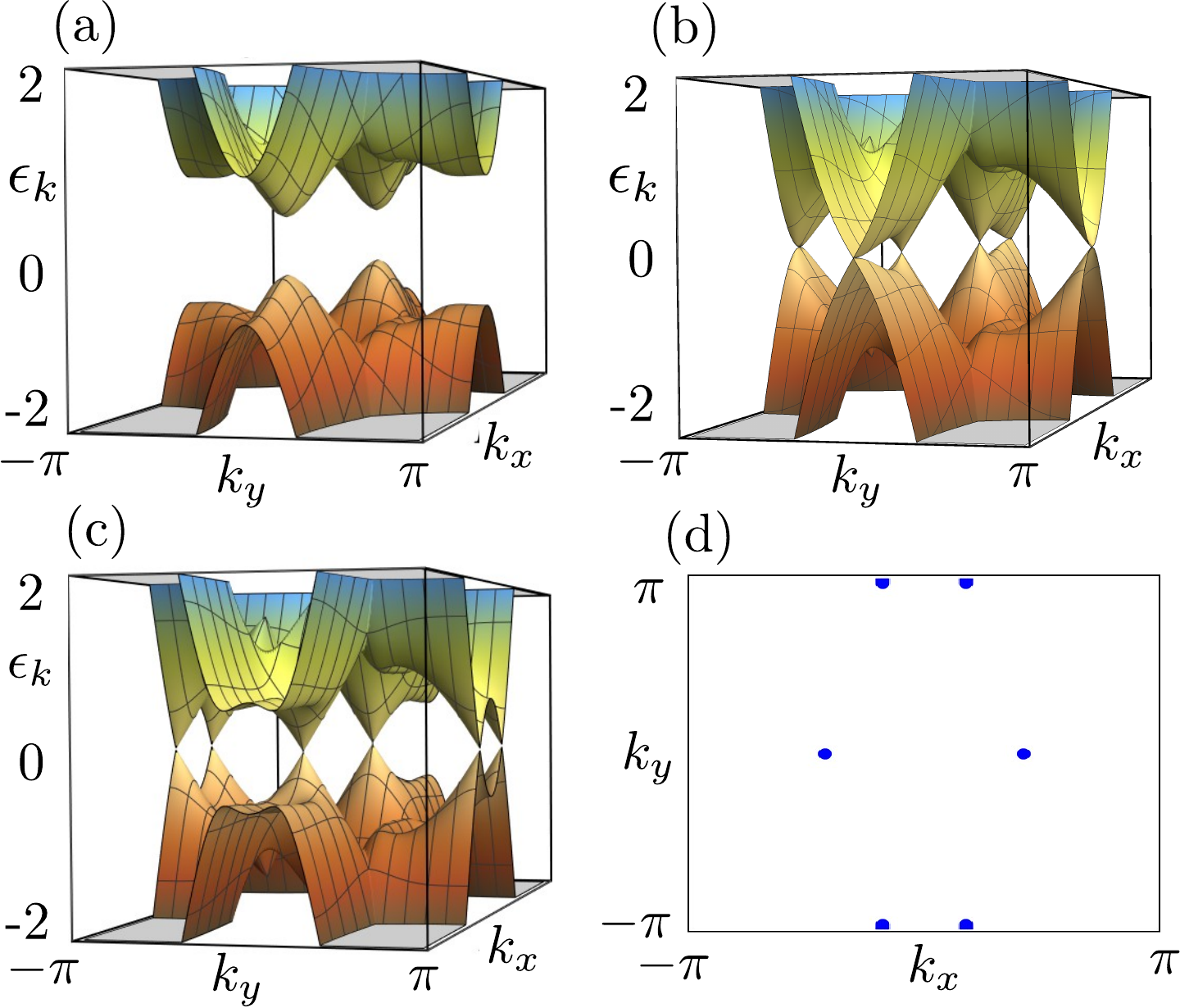}}
\caption{The band structure for a helical $p$-wave superconductor coupled to a $d-$wave altermagnet, as given by Eq.~(\ref{Eq3}). (a) Fully gapped superconducting state for the amplitude of the altermagnet order parameter \( J_A = 0.3t\); 
(b) Topological phase transition at \( J_A =0.5t\); (c) Nodal superconductor at 
\( J_A = 0.65t \); (d) The nodal Fermi surface plot corresponding to the case (c). The values of  parameters are \( \mu = 2t, \Delta_p^h = t \), and \( \Delta_p^c = 0 \), with $t=1$, in Eq.~\eqref{Eq1}.}
\label{Fig1}
\end{figure}

\begin{figure}[t!]
\centering
\subfigure{\includegraphics[width=0.48\textwidth]{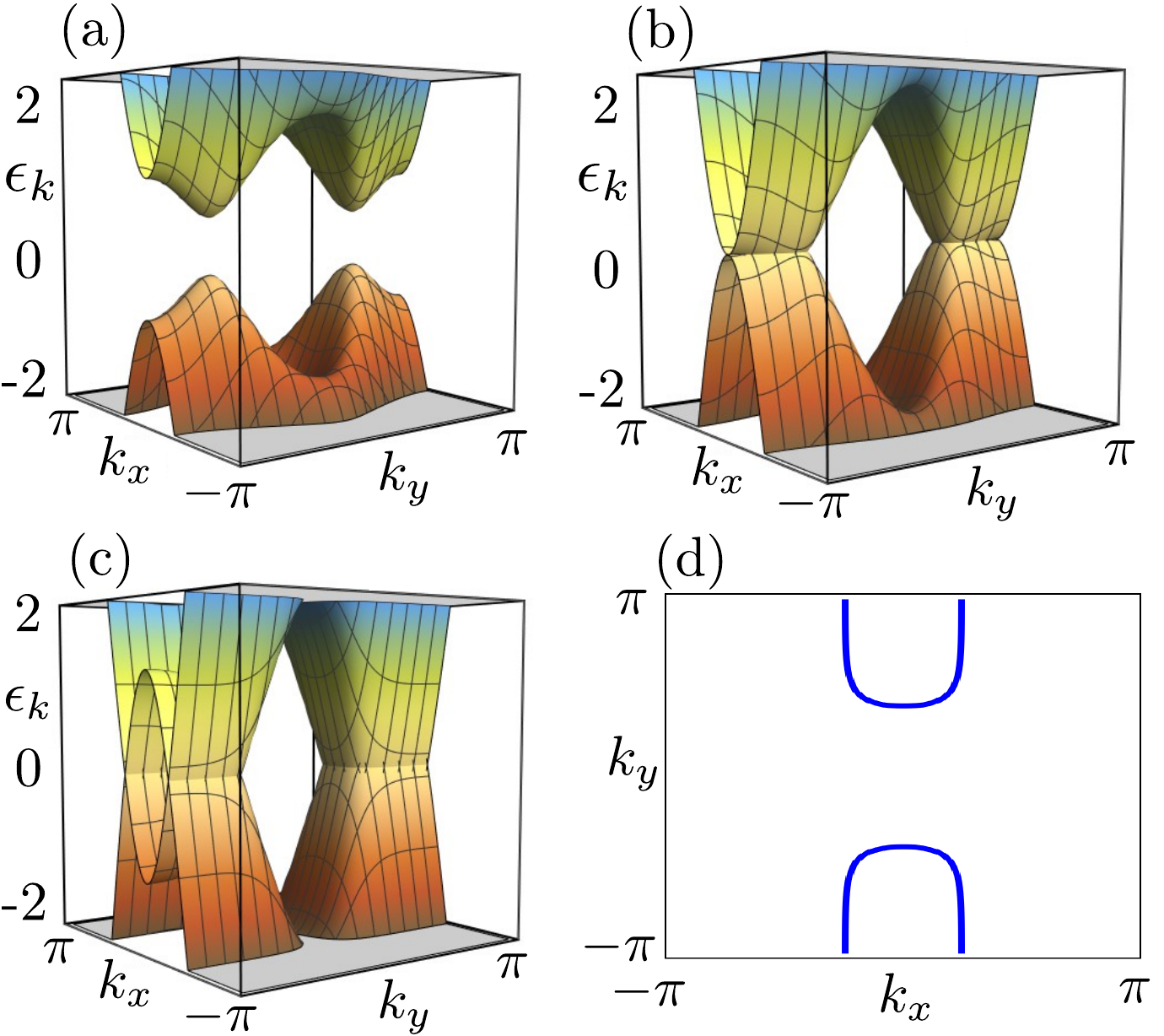}}
\caption{The band structure for a chiral $p$-wave superconductor coupled to an altermagnet, as given by Eq.~(\ref{Eq3}). (a) Fully gapped superconducting state for  \( J_A = 0.3t\); 
(b) Topological phase transition at \( J_A =0.5t\); (c) Nodal-line superconductor at 
\( J_A = 0.8t \); (d) The nodal Fermi surface plot corresponding to the case (c). The values of  parameters are \( \mu = 2t, \Delta_p^c = t \), and \( \Delta_p^h = 0 \), with $t=1$, in Eq.~\eqref{Eq2}.}
\label{Fig2}
\end{figure}

In contrast, for the chiral pairing, the SC term commutes with the altermagnetic OP, and the spectrum simplifies to
\begin{equation}
\epsilon_{\bf k} = r J({\bf k}) + s \sqrt{ \xi_{\bf k}^2 + (\Delta_p^c)^2 (\sin^2 k_x + \sin^2 k_y) },
\label{Eq4}
\end{equation}
with the altermagnetic piece therefore  shifting the spectrum of the Bogoliubov quasiparticles albeit in a momentum-dependent manner, which, in turn, can also yield topologically nontrivial features in the resulting bands. 

Importantly, the interplay between altermagnetism and superconductivity enables topological phase transitions, signaled by the closing and reopening of the quasiparticle gap at high-symmetry points in the Brillouin zone. For the helical case, the critical value of the altermagnetic coupling $J_A$ at which such a transition occurs is found by locating the band-touching point at $X = (\pi, 0)$,
\begin{equation}
J^c_A = \pm \frac{1}{4} |4t - \mu|,
\label{Eq6}
\end{equation}
which identifies a topological phase boundary in parameter space.


We show the band structure [Eq.~(\ref{Eq3})] in Fig.~\ref{Fig1} for a helical $p$-wave superconductor. Notably, a topological phase transition occurs at the amplitude of the altermagnetic OP $J_A = J_A^c$, as given by Eq.~\eqref{Eq6}, separating two distinct topological states. For $J < J_A^c$, the system is always in a gapped, topologically non-trivial phase due to the presence of the helical $p$-wave superconductivity [see Fig.\ref{Fig1}(a)]. In contrast, for $J > J_A^c$, a new topologically non-trivial phase emerges, leading to gapless topological superconductivity [see Fig.\ref{Fig1}(c)], which hosts MFEMs. This new phase features an even number of gapless nodes  [see Fig.\ref{Fig1}(d)], therefore representing a weak topological SC phase~\cite{Fu-Kane-Mele-PRL2007,Moore-Balents-2007}, since it  can be viewed as a stack of one-dimensional superconductors in class D (without TRS), see also Table~\ref{tab:symmetries}, changing the topological invariant at the (gapless) nodal points. 

\begin{figure}[t!]
\centering
\subfigure{\includegraphics[width=0.48\textwidth]{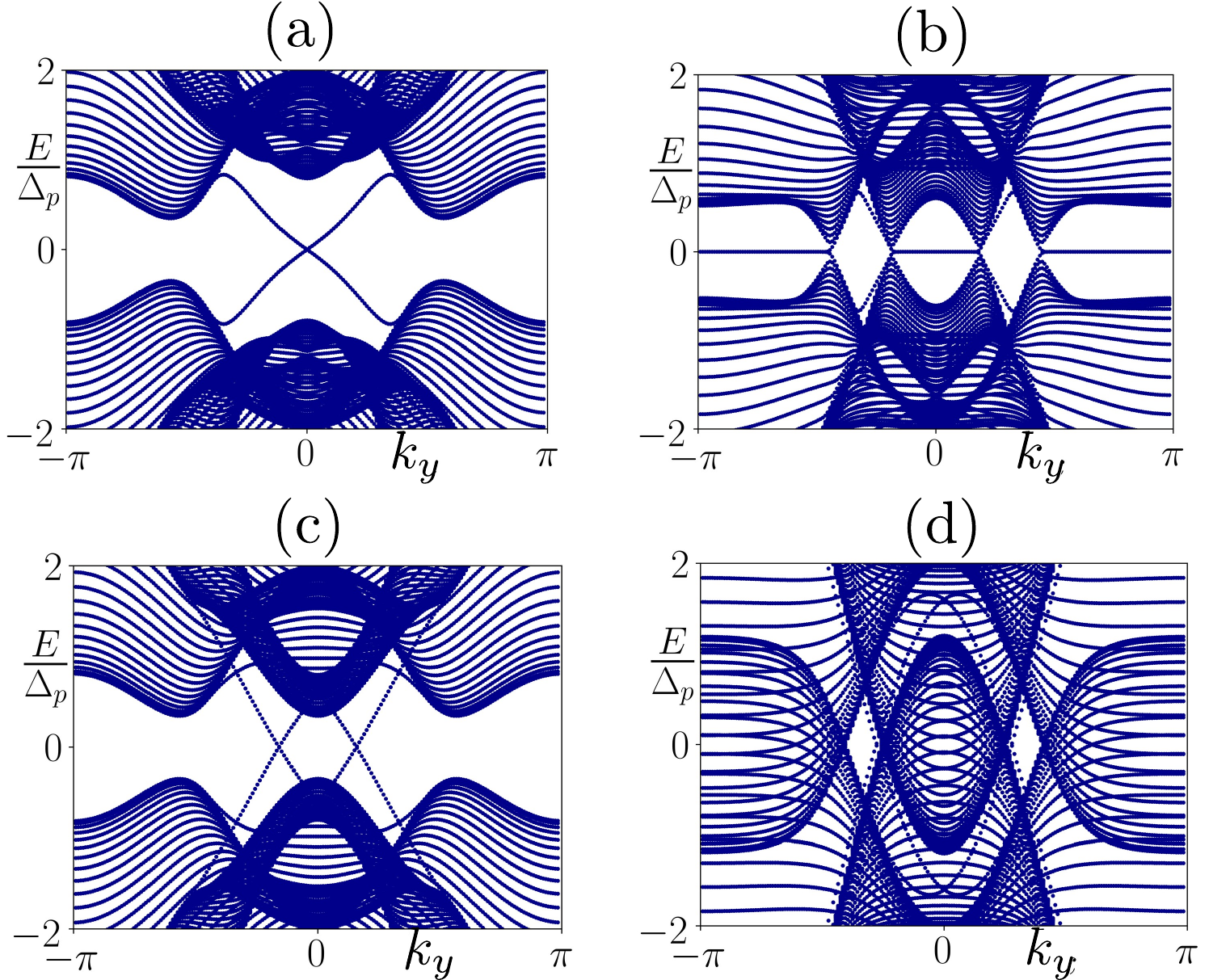}}
\caption{Edge states in the ribbon geometry for a helical and chiral \( p \)-wave superconductor coupled to the altermagnet [Eq.~\eqref{Eq2}]. (a) [(b)] Helical \( p \)-wave superconductor for the altermagnet coupling \( J_A=0.3t \) [\( J_A=0.65t \)]. (c) [(d)] Chiral \( p \)-wave superconductor for the altermagnet coupling  \( J_A=0.3t \)  [\( J_A=0.8t \)], respectively, for a chiral \( p \)-wave superconductor. We employ open [periodic] boundary conditions along \( y \)  [\( x \)] direction and  \( \mu=2t \) is fixed.}
\label{Fig3}
\end{figure}


\begin{figure}[t!]
\centering
\subfigure{\includegraphics[width=0.48\textwidth]{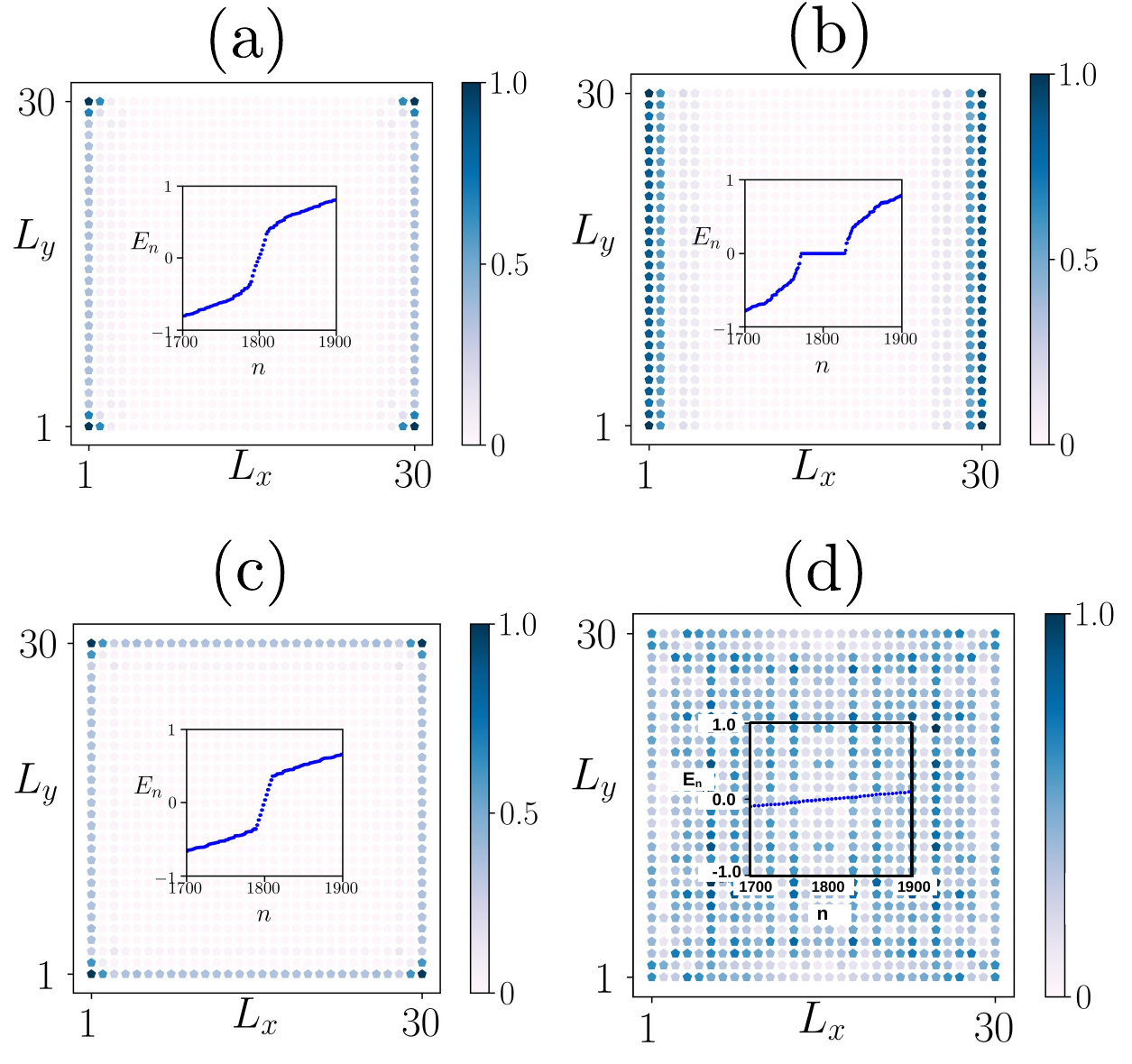}}
\caption{Panels (a) and (b) show local density of states (LDOS) at zero energy ($E=0$) of the  helical $p$-wave superconductor coupled to the altermagnet (Eq.~\eqref{Eq1}) for the values of the  altermagnet amplitude   \( J_A=0.3t \) and \( J_A=0.65t \), respectively. Panels (c) and (d) display LDOS at $E=0$ of the chiral $p$-wave superconductor for the coupling constant \( J_A=0.3t \) and \( J_A=0.8t \), respectively. The inset in each panel shows the spectrum of the closest-to-zero-energy states in the corresponding phase, with the $E=0$ states being of Majorana nature, as corroborated by the characteristic equal distribution of the spectral weight between the particle and hole components of the corresponding wavefunctions. We here set  $\mu=2t$ and $\Delta_p^h=\Delta_p^c=1.0$, and $t=1$.}
\label{Fig4}
\end{figure}

In Fig.~\ref{Fig2}, we display the band structure  [using Eq.~(\ref{Eq4})] for a chiral \( p \)-wave superconductor. We observe a topological phase transition at \( J_A = J_A^c \), where for  \( J_A < J_A^c \), the system is in a gapped, topologically non-trivial phase with broken TRS. On the other hand, for \( J_A > J_A^c \), the system features  a gapless nodal line superconductor [see Fig.~\ref{Fig2}(c)], where the closest to zero energy Bogoliubov quasiparticle bands intersect along a line in the Brillouin zone, with the corresponding form of the Fermi surface displayed in Fig.~\ref{Fig2}(d). Therefore, for a chiral \( p \)-wave superconductor, the \( d \)-wave symmetry of the altermagnet drives the system from a gapped topological phase to a gapless nodal line superconductor. In the following section, we discuss the signatures of these nodal superconductors in terms of the edge states.

\section{Numerical Results}\label{Sec:III}
In this section, we discuss numerical calculations based on the lattice model implemented in  real space given by the Hamiltonian in Eq.~(\ref{Eq1}).  The numerical analysis of edge states and local density of states (LDOS) presented in this section reveals that the interplay between d-wave altermagnetism and unconventional superconductivity gives rise to a range of nontrivial topological phenomena, particularly,  it captures phase transitions and the emergence of Majorana modes across different SC regimes.
\subsection{Edge states in ribbon geometry}\label{subSec:I}


The edge state spectrum in the ribbon geometry is obtained through numerical computations that are carried out by imposing periodic boundary conditions (PBC) along the $y$-direction, implying the conserved $k_y$ momentum along the periodic edge,  and open boundary conditions (OBC) along the $x$-direction.

We first show the spectrum of the edge states for the helical $p$-wave superconductor in the two regimes: $J_A < J_A^c$ [see Fig.\ref{Fig3}(a)] and $J_A > J_A^c$ [see Fig.\ref{Fig3}(b)]. For $J_A < J_A^c$, the system exhibits robust helical edge states, which, however, cease to be protected by TRS  due to the presence of the coupling to the altermagnet, see Table~\ref{tab:symmetries} and Appendix~\ref{app:symmetries}. The loss of TRS protection renders the helical edge states vulnerable to backscattering. Nevertheless, these edge modes can be directly traced back to the helical modes that exist in the absence of any coupling to the altermagnet. Thus, their presence reflects the underlying topological character of the parent helical $p$-wave SC phase, albeit with modified symmetry protection due to the finite $J_A$. 

On the other hand, the system leads to three flat edge modes for $J_A > J_A^c$ [shown in Fig.~\ref{Fig3}(b)], indicating six gapless points in the momentum space, as displayed in Fig.~\ref{Fig1}(d), and therefore the system enters a nodal superconductor phase. This result closely parallels findings in the study of magnetic textures within hybrid superconductor systems \cite{Subhadarshini2024}, featuring  the SC pairing with \(d\)-wave symmetry. Notably, in our context, the \(d\)-wave symmetry of the altermagnet plays a key role in promoting  gapless topological superconductivity, which leads to the emergence of MFEMs in helical \(p\)-wave superconductors.

In a chiral \( p \)-wave superconductor, the ribbon geometry, as expected, supports two chiral edge channels for a small magnitude of the altermagnet  OP, \( J_A < J_A^c \), as displayed in Fig.~\ref{Fig3}(c), with the crossing points moved from the zero momentum since the altermagnetic cooupling commutes with the chiral pairing term. These edge channels are indicative of a  topological SC phase gapped in the bulk that preserves a nontrivial Chern number, and they are localized at opposite edges of the system, reflecting the underlying broken TRS of the chiral pairing. On the other hand, for an overcritical amplitude of the altermagnetic coupling, \( J_A > J_A^c \), the behavior of the system changes dramatically. The edge modes that were previously sharply confined to the boundaries become delocalized, which is  a direct consequence of the system undergoing a topological phase transition into a gapless nodal-line SC phase, wherein the bulk gap closes along isolated lines in momentum space. The nodal-line phase is further confirmed by the gapless contours in the Brillouin zone shown in Fig.~\ref{Fig2}(d). Finally, this transition  is also consequential for the local density of states, which will be discussed next.

\subsection{Local density of states}\label{subSec:II}


To elucidate the presence of the Majorana modes in this system, we now calculate the LDOS, following the approach outlined in Refs.~\cite{Chatterjee2024,Chatterjee2024b,Subhadarshini2024}. The quasiparticle LDOS for the \( i^{\text{th}} \) site at a given energy \( E \) reads 
\begin{equation}
\rho_{i}(E) = \sum_{n\sigma} \left( \lvert u^{n}_{i\sigma} \rvert^{2} + \lvert v^{n}_{i\sigma} \rvert^{2} \right) \delta(E - E_{n}),
\label{Eq7}
\end{equation}
where \( u^{n}_{i\sigma} \) and \( v^{n}_{i\sigma} \) are the electron-like and hole-like quasiparticle amplitudes, respectively, and \( n \) and \( \sigma \) denote the eigenstate and spin indices.

We first consider a helical \( p \)-wave superconductor. In Fig.~\ref{Fig4}, we compute the zero-energy LDOS \( \rho_i(E=0) \) across the two-dimensional plane by implementing open boundary conditions (OBC) along both the \( x \) and \( y \) directions and diagonalizing the lattice Hamiltonian in Eq.~(\ref{Eq2}). For subcritical values of the altermagnetic coupling, \( J_A < J_A^c \), the system hosts dispersive edge modes that are sharply localized along the two boundaries along the $y-$direction, with a negligible LDOS along the $x-$edge, as shown in Fig.~\ref{Fig4}(a), which is also consistent with the low-energy edge theory, as discussed in App.~\ref{sec:edge_helical}. The corresponding eigenvalue spectrum, displayed in the inset, confirms the presence of Majorana zero-energy modes. These dispersive Majorana edge states are a hallmark of the helical topological phase, which remains protected by the particle-hole symmetry (PHS) in the presence of weak altermagnetic perturbations.

As the altermagnetic strength is increased beyond the critical value \( J_A^c \), the system undergoes a topological phase transition into a gapless regime featuring MFEMs. These MFEMs manifest as the states close to zero energy featuring weak dispersion, as illustrated in Fig.~\ref{Fig4}(b). Remarkably, despite the loss of TRS due to the altermagnet, the flat modes remain maximally localized along the same edges, indicating a robust topological character that survives the gap-closing transition. The flatness of these modes in energy signals the collapse of the bulk gap at isolated points in momentum space and the emergence of a weak topological phase composed of symmetry-protected one-dimensional chains.

In the chiral \( p \)-wave phase, the LDOS again reveals a distinct topological character. For \( J_A < J_A^c \), we observe the expected behavior: dispersive chiral Majorana modes are strongly localized along all four edges of the system, consistent with the nonzero Chern number of the chiral SC phase [Fig.~\ref{Fig4}(c)]. For \( J_A > J_A^c \), the LDOS profile changes drastically, as shown in Fig.~\ref{Fig4}(d), where we can observe that the edge localization is lost, and the zero-energy spectral weight is shifted to the bulk. This feature further corroborates  the onset of a nodal-line SC phase, with its signatures previously found  in the band structure and edge state analyses.

\begin{figure}[t!]
\centering
\subfigure{\includegraphics[width=0.5\textwidth]{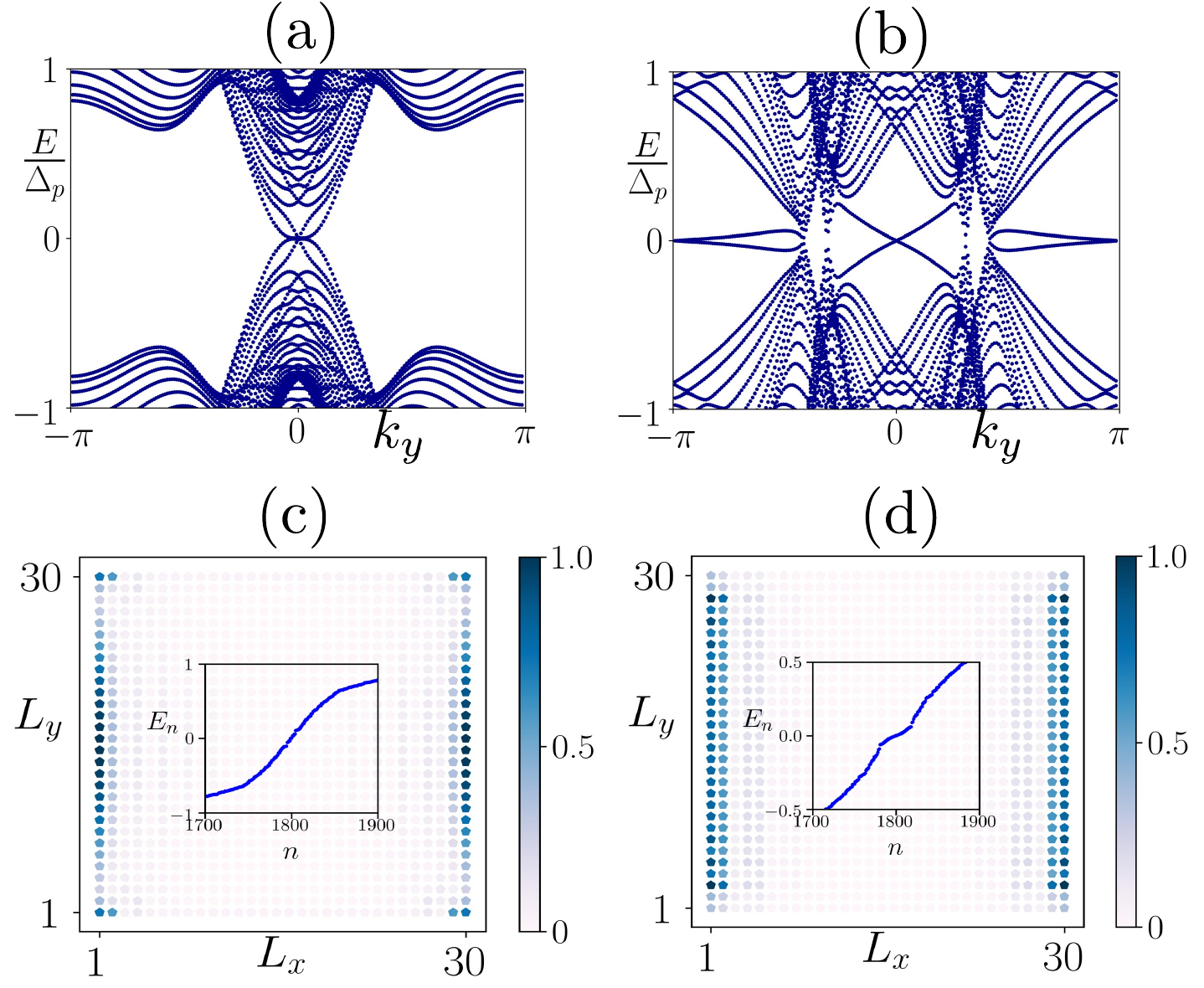}}
\caption{Nearly flat band edge mode driven by altermagnet in the presence of a mixed helical and chiral superconductor. Panels (a) and (b) show ribbon spectrum for a  mixed superconductor with the  amplitude of the altermagnet OP $J_A=0.3t$  ($J_A<J_A^c$) and $J_A=0.65t$  ($J_A>J_A^c$)  respectively, where the critical value is $J_A^c=0.5t$ at which the band structure undergoes the transition. Panels (c) and (d) depict corresponding zero energy local density of states for $J_A<J_A^c$ and $J_A>J_A^c$, respectively. The inset in panels (c) and (d) shows
the spectrum of the closest-to-zero-energy states in the corresponding phase. 
We set the parameters $\mu=2t$ and $\Delta_p^h=\Delta_p^c=t$, with $t=1$, in Eq.~\eqref{Eq1}.}
\label{Fig5}
\end{figure}

\begin{figure*}[]
	\centering
	\includegraphics[width=.9\textwidth]{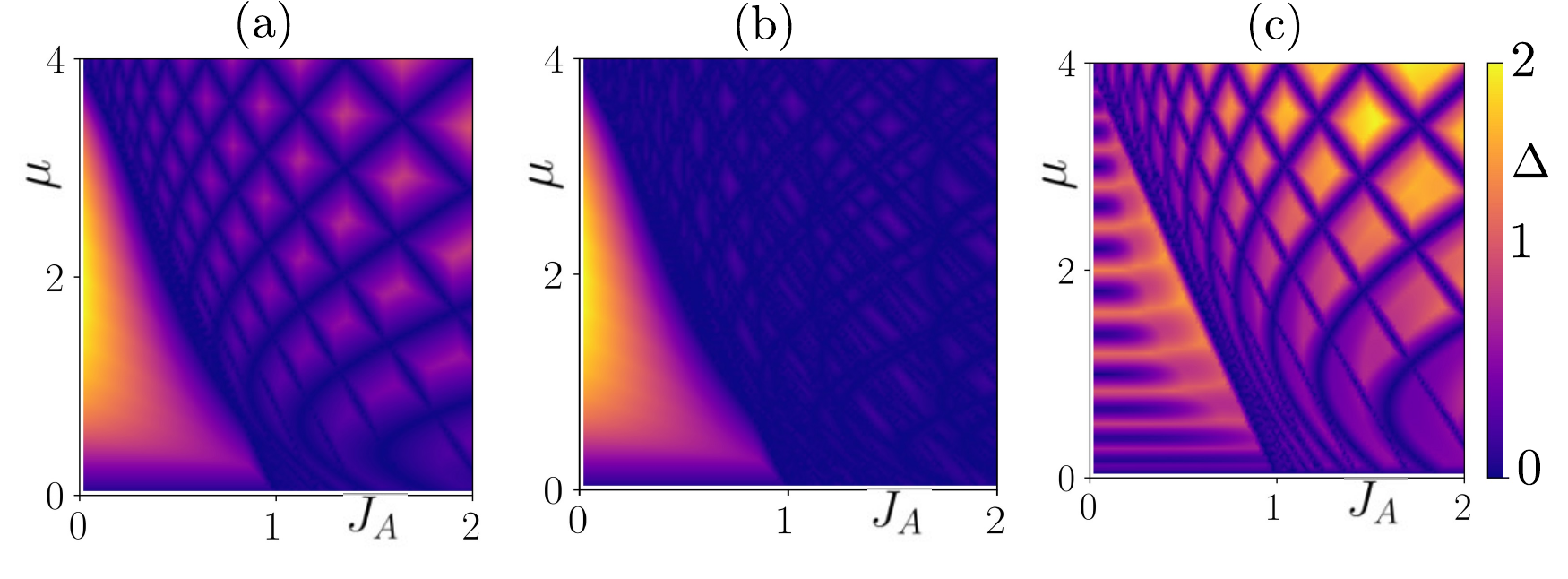}
	\caption{\hspace*{-0.1cm} The bulk gap  ($\Delta$) as a function of the chemical potential ($\mu$) and the amplitude of the altermagnet order parameter  ($J_A$) for helical, chiral, and mixed (helical-chiral) $p$-wave superconductors, shown in panels (a), (b), and (c), respectively. The color bar encodes the size of the gap.  We set the parameters $\Delta_p^h=t$ ($\Delta_p^c = t$) in the helical (chiral) phase, while in the mixed phase   $\Delta_p^h=\Delta_p^c = t$ in Eq.~\eqref{Eq1}, with  $t=1$.   The system size is  $N = 30 \times 30$ sites.}
	\label{Fig6}
\end{figure*}

\subsection{Mixed topological phase}\label{subSec:III}

We now analyze the topological edge states and the LDOS arising from the interplay between $d$-wave altermagnetism and a mixed pairing superconductor with both chiral and helical \( p \)-wave components.

{In Figs.~\ref{Fig5}(a) and \ref{Fig5}(b), we show the ribbon spectrum calculated under periodic boundary conditions (PBC) along the \( y \) direction and open boundary conditions (OBC) along the \( x \) direction. The spectra reveal rich edge-state structures in both the subcritical (\( J_A < J_A^c \)) and overcritical \( J_A > J_A^c \) regimes of the altermagnetic coupling. In the former regime, the system supports multiple dispersive edge modes, which exhibit both linear and quadratic dispersions, as shown in Fig.~\ref{Fig5}(a). These modes emerge as a consequence of  the simultaneous presence of TRS-preserving (helical) and TRS-breaking (chiral) pairing components, which  can hybridize due to the breaking of this symmetry. In the mixed case, both chiral and helical edge states exhibit linear dispersion close to zero edge momentum for $J_A=0$. Upon coupling to the altermagnet, quadratically dispersing modes may emerge alongside the linear ones due to the quadratic momentum dependence of the altermagnet coupling near zero momentum, given its form in Eq.~\eqref{eq:altermagnet}. On the other hand, the linearly dispersing modes in the mixed phase are due to the fully spin polarized form of the chiral SC pairing ($\sim s_0$).  See Fig.\ref{Fig5}(a). 

Notably, when the altermagnet strength exceeds the critical coupling, \( J_A > J_A^c \), the edge spectrum undergoes a qualitative change. As seen in Fig.~\ref{Fig5}(b), the system enters a mixed topological phase characterized by the coexistence of both linearly dispersing and nearly flat edge modes, which can be qualitatively explained as follows. The quadratically dispersing mode in the weak-coupling regime due to an enhanced $E=0$ DOS and the lack of topological protection is rendered unstable  at strong coupling. Since  the chosen $d-$wave altermagnetic OP  does not open a full gap at the edges [the $y-$edge remains gapless, as implied by Eq.~\eqref{eq:y-edge-kappa}, and explicitly shown in App.~\eqref{sec:edge_helical}],  to minimize the kinetic energy , the spectral density of the quadratic band is transferred to low energies while forming an almost flat band.    Such nearly flat modes feature localized, zero-energy MFEMs. On the other hand, the linearly dispersive branches at the center of the Brillouin zone remain gapless at strong coupling due to their chiral nature and the fact that the chosen altermagnetic coupling does not open an edge gap in the chiral case.

The presence of both flat and dispersive edge states in this mixed regime is further supported by the $E = 0$ LDOS profiles, shown in Figs.~\ref{Fig5}(c) and \ref{Fig5}(d). For both subcritical and supercritical values of \( J_A \), the zero-energy states are  localized only at the $y-$edge of the system. The inset plots display the low-energy eigenvalue spectrum, confirming the presence of quasi-zero-energy states.

\section{gap profile}\label{Sec:IV}
To gain further insight in the phase diagram of the model, in this section, we analyze the variation of the bulk gap profile by employing PBC along both the $x$ and $y$ directions as a characteristic signature of the topological phase transition.

We calculate the topological invariant for  the purely chiral and helical phases in the Hamiltonian~\eqref{Eq2} for the weakly coupled regime $J_A<J_A^c$. The Chern number can be expressed as~\cite{Mascot2021,Teemu2016}
\begin{equation}
C = \frac{1}{2\pi i} \smashoperator{\int_{\rm BZ}}\! d^2k\, \text{Tr} \Big( P^- \big[ \partial_{k_x} P^-, \partial_{k_y} P^- \big] \Big)
\label{Eq9}
\end{equation}
in terms of the projector to the two filled bands $P^- = P_1^- + P_2^-$, where
\begin{equation}
\begin{split}
P_1^- = \frac{(E_1-H)(E_2-H)(E_{-2}-H)}{(E_1-E_{-1})(E_2-E_{-1})(E_{-2}-E_{-1})},
\\
P_2^- = \frac{(E_1-H)(E_2-H)(E_{-1}-H)}{(E_1-E_{-2})(E_2-E_{-2})(E_{-1}-E_{-2})}.
\end{split}
\label{Eq10}
\end{equation}
Here, $E_{-1}$ and $E_{-2}$ are the energies of the two filled bands, and $E_1$ and $E_2$ are the energies of the unoccupied bands. In both cases, the Chern number turns out to be equal to one, therefore corroborating the topological nature of these phases. To distinguish between the gapped and gapless topological SC phases, we calculate the bandgap and analyze the form of the edge states.

As an indicator of the topological phase transition, in Fig.~\ref{Fig6}, we show the maximum value of the gap, \(\Delta = |E_2 - E_1|\),  between closest to zero energy Bogoliubov quasiparticle bands, where \(E_1\) (\(E_2\))  represents the maximum (minimum) of the energy of the negative (positive) energy band.  Fig.~\ref{Fig6}(a) displays the gap, \(\Delta\), as a function of the altermagnet strength \(J_A\) and chemical potential \(\mu\) for the helical \(p\)-wave superconductor. For a fixed (nonzero) value of \(\mu\), we observe that this gapped topologically non-trivial  phase remains robust  in the regime \(J_A < J_A^c\), where $J_A^c$ is the critical coupling for the phase transition at the zero chemical potential. On the other hand, for a strong altermagnet amplitude,  \(J_A > J_A^c\), we observe multiple topological phase transitions in terms  of the chemical potential, with the bulk gap closing and reopening. Each gap closing point corresponds to a gapless topological SC phase hosting MFEMs, as depicted in Fig.~\ref{Fig3}(b).

In Fig.~\ref{Fig6}(b), we display the outcome of  an analogous analysis  for a chiral $p$-wave superconductor. For a fixed value of $\mu$, we observe a topological phase transition from a chiral gapped SC phase to a gapless nodal-line SC phase, see also Fig.~\ref{Fig3}(d). Finally, in Fig.~\ref{Fig6}(c), we show the maximum gap $\Delta$ in the $\mu-J_A$ plane for a mixed $p$-wave superconductor. Interestingly, in this case, we also find multiple topological phase transitions between gapped and gapless phases, which is a clear indication of a mixed topological phase that hosts MFEMs and dispersive edge states, shown in Fig.~\ref{Fig5}(b).
\section{Summary and discussion}\label{Sec:V}
In summary, we investigated the effect of a $d$-wave symmetric altermagnet texture  on a 2D helical and chiral $p$-wave superconductor. For a helical $p$-wave superconductor, altermagnetism drives the system from a helical gapped topological superconductor to a gapless topological superconductor hosting MFEMs. However, for a chiral $p$-wave superconductor, such a coupling yields a topological phase transition into a gapless nodal line superconductor. Most interestingly, for a mixed (helical-chiral) $p$-wave superconductor, altermagnetism induces a new hybrid topological phase where dispersive and flat Majorana edge modes coexist. 

In this work, our primary objective was to identify signatures of flat and mixed (dispersive + flat) edge modes in an altermagnet-superconductor hybrid system. Based on the $d-$wave nature of the  altermagnet OP, we expect that the edge states can be controlled by  orientation of the altermagnet, characterized by the azimuthal and polar angles,  $\theta$ and $\phi$, respectively.  To this end,  in our model, the altermagnet couples to spin along the $x$-direction, i.e., $\theta = \pi/2$, $\phi = 0$, which results in a coupling term that neither commutes nor anticommutes with the Hamiltonian of  parent $p$-wave topological superconductor. This specific form leads to selective gapping of the corresponding edge-state Hamiltonian. Consequently, the edge states are localized  along the $y$-edges, corresponding to $L_x = 1$ and $L_x = 30$, in the mixed (helical case), as shown in Figs.~\ref{Fig5}(c) and (d) [\ref{Fig4}(a) and (b)].  Notably, because the superconductor-altermagnet coupling along the $x$-direction does not constitute a mass term for the parent Hamiltonian, the emergence of higher-order topological phases is precluded in our setup. This should be contrasted with the situation in Ref.~\cite{Li2024}, where a different form of the coupling enables a sign-changing Dirac mass, thereby allowing for the realization of a higher-order topological SC phase in a parent helical $p-$wave superconductor.

Our findings, particularly, the predicted  gapless topological phase  and MFEMs may be relevant in light of a recent experimental proposal~\cite{Wiesendanger2024}, where a gapless topological SC phase hosting MFEMs in a magnet-superconductor heterostructure composed of Fe/Ta(110) has been identified. Here, we propose an alternative method to engineer gapless topological superconductivity hosting MFEMs in such a hybrid system. Specifically, the potential  experimental setup realizing our scenario involves $\rm{Ru}\rm{O}_2$, a leading candidate for altermagnetism with $d$-wave magnetic order~\cite{Sinova2022b} (see, however, Ref.~\cite{plouff2024revisiting}), fabricated  on top of a topological $p$-wave superconductor, such as  $\rm{U}\rm{Te}_2$~\cite{Jiao2020}. This heterostructure could serve as a promising candidate for engineering gapless topological superconductivity hosting MFEMs. Although this remains experimentally challenging due to the limited availability of topological 
$p-$wave SC materials, our study  opens new avenues for engineering gapless, and, particularly, hybrid topological SC states, wherein Majorana dispersive and flat edge modes coexist within an unconventional  SC platform featuring  zero net magnetization. Finally, our work should motivate further studies on the characterization of the transition between  gapped and nodal-point or nodal-line superconductor, in particular the topological and thermodynamic features of such transitions~\cite{Yerin_2022,Yerin-SciPost-2022}. 

\subsection*{Acknowledgments}
PC acknowledges Arijit Saha, Kenji Fukushima, and Takashi Oka for fruitful discussions and support.
V. J. acknowledges the  support of  Fondecyt (Chile) Grant  No. 1230933 and the Swedish Research Council Grant No. VR 2019-04735.

\subsection*{Data Availability}

The data that support the findings of this study are available from the authors upon reasonable request.

\appendix

\section{Symmetries of the Bogoliubov-de Gennes Hamiltonian in Eq.~\eqref{Eq2}}\label{app:symmetries}

In this Appendix, we consider the discrete symmetries of the BdG Hamiltonian introduced in Eq.~(\ref{Eq2}), anti-unitary {particle-hole symmetry (PHS)}, {time-reversal symmetry (TRS)}, and {chiral symmetry (CS)} (or unitary PHS). The results of this analysis are summarized in Table~\ref{tab:symmetries}.

\subsection{Anti-unitary particle-hole symmetry } 

The BdG Hamiltonian features PHS by construction, with the corresponding  anti-unitary operator $\mathcal{P}$ acting  on the Nambu spinor as
\[
\mathcal{P} = \tau_x s_z \mathcal{K},
\]
where $\mathcal{K}$ denotes complex conjugation. The symmetry condition for the Hamiltonian $ h_{\bf k}$ reads as
\[
\mathcal{P} h_{\bf k} \mathcal{P}^{-1} = -h_{-{\bf k}}.
\]
Using the structure of $h_{\bf k}$ in Eq.~(\ref{Eq2}), we find that terms transform  under $\mathcal{P}$ as follows: 
\begin{enumerate}[(a)]
    \item $\xi_{\bf k} \tau_z s_0 \rightarrow -\xi_{-{\bf k}} \tau_z s_0$ (since $\xi_{\bf k} = \xi_{-{\bf k}}$),
    \item $J({\bf k}) \tau_0 s_x \rightarrow -J({\bf k}) \tau_0 s_x$ (with $J({\bf k}) = J(-{\bf k})$),
    \item pairing terms  $\sim\sin k_{x,y}$ are odd in momentum multiplied by the matrices that commute with the particle-hole operator, $\mathcal{P} = \tau_x s_z \mathcal{K}$, for both helical and chiral SC OPs.
\end{enumerate}
Hence, the  Hamiltonian in Eq.~\eqref{Eq2} possesses anti-unitary  PHS with $\mathcal{P}^2 = +1$. 
\subsection{Time-Reversal Symmetry (TRS).} 
The time-reversal operator for spin-$1/2$ systems is given by
\[
\mathcal{T} = \tau_0 (i s_y) \mathcal{K},
\]
with the TRS condition:
\[
\mathcal{T} h_{\bf k} \mathcal{T}^{-1} = h_{-{\bf k}}.
\]

\textit{(a) Chiral $p$-wave case ($\Delta_p^h = 0$):}
\\

The chiral pairing term 
\[
\Delta_p^c (\tau_x s_x \sin k_x - \tau_y s_x \sin k_y)
\]
is \textit{odd} under time reversal:
\begin{align*}
\mathcal{T} \tau_x s_x \mathcal{T}^{-1} &= -\tau_x s_x, \\
\mathcal{T} \tau_y s_x \mathcal{T}^{-1} &= -\tau_y s_x, \\
\sin k_{x,y} &\rightarrow -\sin k_{x,y}.
\end{align*}

The altermagnetic term $J({\bf k}) \tau_0 s_x$ also breaks TRS:
\[
\mathcal{T} \tau_0 s_x \mathcal{T}^{-1} =- \tau_0 s_x, \quad J({\bf k}) = J(-{\bf k}).
\]
Therefore, the chiral $p$-wave superconductor {explicitly breaks TRS}, even without the altermagnetic term.
\\

\textit{(b) Helical $p$-wave case ($\Delta_p^c = 0$):}
\\

The helical pairing term is
\[
\Delta_p^h (\tau_y s_0 \sin k_x - \tau_x s_z \sin k_y).
\]
Under TRS:
\begin{align*}
\mathcal{T} \tau_y s_0 \mathcal{T}^{-1} &= -\tau_y s_0, \\
\mathcal{T} \tau_x s_z \mathcal{T}^{-1} &= \tau_x s_z, \\
\sin k_{x,y} &\rightarrow -\sin k_{x,y}.
\end{align*}
Thus, both terms in the helical pairing are \textit{even} under TRS. However, the altermagnetic term $J({\bf k}) \tau_0 s_x$ again breaks TRS, as in the chiral case. Therefore, while the helical superconductor alone {preserves TRS}, the inclusion of altermagnetism leads to {TRS breaking}.
\subsection{Chiral Symmetry (CS).} 
Chiral or unitary particle-hole symmetry is defined as the combination of anti-unitary PHS and TRS:
\[
\mathcal{C} = \mathcal{P} \mathcal{T} = \tau_x s_z \cdot \tau_0 i s_y = \tau_x s_x.
\]

The symmetry condition is:
\[
\mathcal{C} h_{\bf k} \mathcal{C}^{-1} = -h_{\bf k}.
\]
However, since the Hamiltonian with altermagnetism {breaks TRS}, chiral symmetry is {absent} in both the helical and chiral cases.

In summary, the BdG Hamiltonians with altermagnetic order, supporting both chiral and helical SC OPs,  belong to symmetry class {D} in the tenfold classification of topological insulators and superconductors. This class supports a $\mathbb{Z}$ topological classification in two spatial dimensions~\cite{ryu2010topological} and therefore  implies the possible existence of topologically protected chiral Majorana edge modes.

\section{Edge-state analysis: Purely helical pairing}\label{sec:edge_helical}

To provide an analytic understanding of the edge-state structure in the presence of helical $p$-wave SC OP, we consider the low-energy expansion of the Bogoliubov--de Gennes (BdG) Hamiltonian in Eq.~\eqref{Eq2} with the chiral pairing $\Delta_p^c=0$. We analyze the edge states by considering a continuum limit near the $X-$point in the Brillouin zone at the momentum, $\vb{K}_X=(\pi,0)$,  and imposing semi-infinite boundary conditions, therefore considering:  (1) the $y-$edge, with the $x-$direction open and the $k_y$ momentum conserved;  (2) the $x-$edge, with the $y-$direction open and the $k_x$ momentum conserved. 

To this end, we first expand the Hamiltonian~\eqref{Eq2} about the $X-$point in the Brillouin zone to obtain,
\begin{align}\label{eq:edge-hel}
h_X(\vb{k}) &= \left[ -t(k_x^2 -k_y^2) +4t- \mu \right] \tau_z s_0\nonumber\\
&+[-4J_A + J_A(k_x^2 + k_y^2)] \tau_0 s_x - \Delta_p^h \left( k_x \tau_y s_0  +k_y \tau_x s_z  \right).
\end{align}
Here, the $\vb{k}$ is a small deviation of the momentum from the $X-$point, $|\vb{k}|\ll|\vb{K}_X|$. 

\subsection{Edge parallel to the  $y$-direction}

We now replace $k_x \to -i\partial_x$, keeping $k_y$ as a good quantum number in Eq.~\eqref{eq:edge-hel}
\begin{align}
H_{y\text{-edge}} &= \left[ t(k_y^2 + \partial_x^2) +4t- \mu \right] \tau_z s_0 + J_A(-4-\partial_x^2 + k_y^2) \tau_0 s_x \notag \\
&+ \Delta_p^h \left( i \partial_x  \tau_y s_0 - k_y \tau_x s_z  \right).
\end{align}
We look for the zero energy  solutions of the form $\psi(x) = e^{-\kappa x} \chi$, such that ${\rm Re}(\kappa)>0$ ensuring normalizability of the mode, which subject to open boundary conditions at $x=0$, $\psi(x=0)=0$. To find such a zero mode we take   $k_y=0$ and  the low-energy effective edge Hamiltonian then takes reduces to 
\begin{align}\label{eq:y-edge-kappa}
H_{y\text{-edge}} &= \left[ t \kappa^2 + 4t - \mu \right] \tau_z s_0 - J_A (4 + \kappa^2) \tau_0 s_x + i\Delta_p^h \kappa \tau_y s_0.
\end{align}
The inverse localization length of the  zero energy Majorana modes, $\kappa$, is obtained from the condition of the vanishing determinant of this matrix. We then find 
\begin{align}
\mathrm{det}\,H_{y\text{-edge}} &=
(J_A^2 - t^2)\kappa^4 
+ (8 J_A^2 - 8 t^2 + \Delta_h^2 + 2 t \mu)\kappa^2 \nonumber \\[6pt]
&\quad + (16 J_A^2 - 16 t^2 + 8 t \mu - \mu^2)
\end{align}
which we will not analyze in detail. We just point out that for the values of the parameters in Figs,~\ref{Fig3}(a) and~\ref{Fig4}(a) we find a pair of localized zero modes, which indeed yield a pair of counterpropagating edge modes obtained by the diagonalization of the Hamiltonian~\eqref{Eq2} in a semi-infinite geometry, with the spectrum shown in Fig.~\ref{Fig3}(a). This behavior is also expected based on the form of the edge Hamiltonian~\eqref{eq:y-edge-kappa} where the altermagnetic term commutes with the pairing term, which is the effective kinetic term in the edge Dirac Hamiltonian. Therefore, such a term  is expected not to gap them out, and these modes being topological should remain at zero energy.

\subsection{Edge parallel to the  $x$-direction}

Next, we consider a $x$-edge geometry and perform a similar expansion:
\begin{align}\label{eq:x-edge}
H_{x\text{-edge}} &= \left[ -t(k_x^2 + \partial_y^2)+4t - \mu \right] \tau_z s_0 + J_A(-4-\partial_y^2 + k_x^2) \tau_0 s_x \notag \\
&\quad - \Delta_p^h \left(k_x \tau_y s_0  - i \tau_x s_z \partial_y \right).
\end{align}
Using the ansatz $\psi(y) = e^{-\kappa y} \chi$, the projected effective Hamiltonian for the $x$-edge for the zero mode becomes
\begin{align}\label{eq:x-edge-1}
&H_{x\text{-edge}}=\left(-t \kappa^2 + 4 t - \mu \right) \, \tau_z s_0 
+ \left(-4 J_A - J_A \kappa^2 \right) \, \tau_0 s_x\nonumber\\
&- i\, \Delta_h \kappa \, \tau_x s_z,
\end{align}
which is obtained after setting $k_x=0$ in  Eq.~\eqref{eq:x-edge}. The corresponding condition for the zero modes is obtained by setting the determinant of this Hamiltonian to zero, which explicitly reads 
\begin{align}
&\ {\det} (H_{x\text{-edge}})= J_A^4 (4 + \kappa^2)^4 \notag \\
&\ + \left[
    t^2 (-4 + \kappa^2)^2 + 2t(-4 + \kappa^2) \mu + \mu^2 - \Delta_h^2 \kappa^2
   \right]^2 \notag \\
&\ - 2 J_A^2 (4 + \kappa^2)^2 \left[
    \Delta_h^2 \kappa^2
    + t^2 (-4 + \kappa^2)^2
    + 2t(-4 + \kappa^2) \mu
    + \mu^2
   \right]
\end{align}
For the regime of parameters in Figs.~\ref{Fig3}(a) and \ref{Fig4}(a), the above expression yields four distinct positive purely real values for the inverse localization length $\kappa$ (besides four negative ones, which correspond to non-normalizable modes). However, these modes are incompatible with the open boundary condition at $y=0$, and therefore no zero modes exist in this case. This outcome may be expected also based on the fact that the  altermagnetic term anticommutes with the pairing term in the effective edge Hamiltonian~\eqref{eq:x-edge-1}, and therefore it  is expected to gap out the edge modes. 

These results are in agreement with our numerical findings displayed  in Fig.~\ref{Fig3}(a) and~\ref{Fig4}(a).

The case of the edge states in the  chiral SC  can be analyzed in a similar manner but we do not delve into this case here.

\bibliography{ArXiv-AM-pSC}

\end{document}